# Converging the Neutrino mass limit: An experimental approach

*Athar Ahmad\*, Naila Islam\**

*\*Department of Physics, Aligarh Muslim University*

*Abstract:* Neutrinos are the second most abundant particle in the universe. Since the last 50 years, Neutrino physics has been a source of limelight in modern physics because of the incredible characteristics of this elusive particle. According to the basic postulates of the Standard Model of particle physics neutrinos were assumed to be massless, but the results inferred from the experiments conducted in the past few years in the field of neutrino physics have concluded that neutrinos have a finite mass. In this paper we have discussed the various experiments conducted in the past which have tried to set an experimental limit on the absolute neutrino masses. Further, the future experiments which are expected to bring these limits to sub eV regime are also discussed.

The Standard model of particle physics is considered to be one of the most important triumphs of modern physics. The theoretical framework developed in the 70's, is used to provide the theory behind the three fundamental forces (Weak Force, Strong Nuclear Force and Electromagnetic Force, excluding Gravitational Force) and the elementary particles constituting the universe. There are various factors which support the validity of this framework. One such factor is the assumption that neutrinos are massless.

This assumption was first opposed by Bruno Pontecurvo in 1957 when he proposed the idea of matter antimatter oscillations analogous to the oscillations in mixed neutral particles (specifically neutral kaon mixing) [1]. This assumption was further investigated by Z. Maki, M. Nakagawa and S. Sakata in 1962 to postulate the neutrino flavour mixing angle thus defining specific mass eigenstates for neutrinos which led to the concept of neutrino oscillations [2]. After Pontecurvo suggested some experimental setups to detect $\nu \rightleftharpoons \bar{\nu}$ and $\nu_e \rightleftharpoons \nu_\mu$ oscillations in 1967 [3], the Homestake experiment headed by Raymond Davis Jr. and John N. Bahcall successfully found a deficit in the detection of solar neutrinos as only 1/3 of the expected $\nu_e$ neutrinos were detected in the experiment [4]. This laid the foundation to the famous solar neutrino problem. The problem was justified by particle physicists as an error in the solar model or a possibility of physics beyond the standard model in the form of neutrino oscillations. After decades of experimental research, the solution to this problem was experimentally presented by the SNO observatory. Their results were in close agreement with the predicted value of 34% of the $v_e$ neutrinos i.e. due to the neutrino oscillations only 34% of the solar $v_e$ were detected [5,6]. The experimental confirmation of neutrino oscillations thus justified the existence of non-zero neutrino rest mass.

Several experiments have been conducted in the past to determine a mass limit on the absolute neutrino masses. The three most reliable approaches which can lead us to the absolute neutrino mass scale are:

*Cosmological sources*: Using the observed neutrino density the standard cosmological constant gives us an upper bound on the sum of the neutrino masses ($\Sigma m_v$). In the cosmological approach, we measure the matter density by reviewing the signals like the Cosmic Microwave Background where we study the distribution of matter by scanning the nearby galaxies to get a structure of the universe at different scales. Comparing the observational data with the computer simulations of the same and using this comparative data in the $\Lambda CDM$ model we can calculate an upper bound on the sum of neutrino mass ($\Sigma m_v$). Using the Planck 2018 data of baryon acoustic oscillation from LSS, the upper bound of $\Sigma m_v$ was estimated to be less than 0.12 eV. Although this prediction is completely model dependent as nearly 95% of the energy distribution of the $\Lambda CDM$ model which contains the dark matter and dark energy is still unknown.

*Neutrino less Double β Decay (0νββ)*: A $0\nu\beta\beta$ decay involves, $(Z, A) \rightarrow (Z - 2, A) + 2e^-$ i.e. two $e^-(e^+)$ are emitted simultaneously as a result of two $\beta$ decays in a nucleus at the same time where a Majorana neutrino is exchanged between the two reactions (Fig. 1). This reaction leads to lepton number violation and has not yet been experimentally observed. $0\nu\beta\beta$ is only possible in the case when a Majorana neutrino($\bar{v} = v$) mediates the process. Hence, by experimentally calculating the lifetime ($\tau$) of the reaction, we can obtain the value of absolute neutrino mass as,

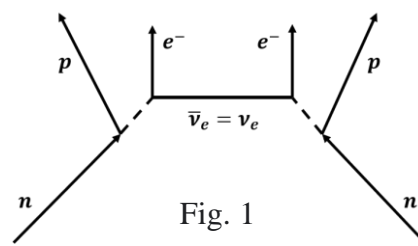

Fig. 1

$$\Gamma_{0\nu\beta\beta}\left(=\frac{1}{\tau}\right) \propto \left|\Sigma U_{ei}^2 m(\nu_i)\right|^2 := m_{ee}^2$$

where, $\Gamma_{0\nu\beta\beta}$ is the decay rate of the reaction

$U_{ei}$ is the neutrino mixing matrix

$m(\nu_i)$ is the neutrino mass

$m_{ee}$ is the sum of neutrino masses

*Direct Neutrino Mass Determination:* Instead of a model-based approach, we use direct kinematic calculation based on the energy momentum relation from special theory of relativity, i.e. $E^2 = p^2 c^2 + m_\nu^2 c^4$ where $m_\nu^2$ is the observable. There are two methods to perform Direct Neutrino Mass Determination. We can use the time of flight measurement of neutrinos emitted from a catastrophic astronomical event like a core collapse supernova. An alternative approach is by investigating the kinematics of weak decays in laboratory. Specifically, by analysing the end point region of β decay spectrum we can calculate the neutrino mass. This is currently the most preferable method to measure neutrino mass as it is model independent and extremely sensitive in the sub eV regions.

Analysing the *cosmological sources* like Wilkinson Microwave Anisotropy Probe (WMAP) 3-year data, Sloan Digital Sky Survey (SDSS) measurement of the baryon acoustic peak, the Type-1a supernova from Supernova Legacy Survey (SNLS) we have estimated an upper bound on the sum of neutrino masses ($\Sigma m_\nu$) to be $\leq 0.62\ meV$ (95% $C.L.$). Using the Lyman-$\alpha$ analysis the bound on the $\Sigma m_\nu$ has reduced to a scale of 0.2-0.4 eV (95% C.L.) [7].

The evidence of *Neutrino less Double β Decay (0νββ)* had been claimed by the Heidelberg-Moscow collaboration (HDM) in their first run in 2001 with an alleged life time ($T_{\beta\beta}^{0\nu}$) of >2*10$^{25}$ years [8]. Later Germanium Detector Array (GERDA) performed the same experiment in two runs but were unable to detect a signal confirming $0\nu\beta\beta$ and hence estimated an upper bound of $T_{\beta\beta}^{0\nu} > 5.3*10^{25}$ years (90% C.L.) [9]. Further experiments like Enriched Xenon Observatory (EXO-200) and KamLAND-Zen (Kamioka Liquid Scintillator Antineutrino Detector-Zen) have used Xenon sources to search for $0\nu\beta\beta$ and have obtained an upper bound of 1.1*10$^{25}$ years and 2.6*10$^{25}$ years with a 90% C.L. respectively [10]. The

EXO-200 experiment has set an upper bound of 133 meV to 186 meV to the squared sum of Majorana neutrino masses [10].

The *Direct Neutrino Mass Determination* by the time-of-flight approach was first used by Kamiokande-II, Irvine–Michigan–Brookhaven (IMB) collaboration and Baksan experiment to detect high energy neutrinos from an event of core collapse supernovae (SN1987a) in the Large Magellanic Cloud (LMC) on 23 February, 1987. Although it was expected that SN1987a would emit neutrinos/anti-neutrinos of all flavours, but due to high threshold or too low cross section only electron anti-neutrinos($\bar{v_e}$) were detected by the inverse β decay process($\bar{v_e} + p \rightarrow n + e^+$). The neutrino burst event was detected simultaneously by the three respective detectors and by further calculations using the time-of-flight method, an upper bound on the rest mass of electron type anti-neutrino($\bar{v_e}$) was estimated to be 5.7 eV and 5.8 eV using two different supernova models [11][12].

The most convenient way of measuring neutrino mass is by investigating the kinematics of β decays in the laboratory. The first experiment of this kind was reported in 1949 when S.C. Curran and his team used tritium as a source to determine a bound of 1 keV on $m^2(v_e)$[13]. Another claim was reported in the 80's, by the Institute of Theoretical and Experimental Physics (ITEP), Moscow when they used a thin film of tritiated valine as a β source with a Tretyakov spectrometer to predict an upper bound of 30eV on $v_e$ mass [14,15]. But their claim was soon disapproved by the University of Zürich and Los Alamos National Laboratory (LANL) when they used a more advanced tritium source in the form of a solid source of tritium implanted into carbon and a gaseous molecular tritium source respectively along with the same Tretyakov spectrometer. Using the method of electron capture($\beta^+ decay$), P.T. Springer (and team, 1987) and S. Yasumi (and team, 1994) performed experiments where they used [163]Ho as a source to obtain an upper limit of 225 eV (95% C.L.) and 490 eV (68% C.L.) on the average $v_e$ mass [16,17]. Another prediction of the $v_e$ mass limit was obtained by M. Jung (and team, 1992) where they used [163]Dy[66+] source to perform the first bound state $\beta^-$ decay. Their results predicted a maximum value of 410 eV (68% C.L.) for $v_e$ mass [18]. In the 90's, further results came using tritium as β decay sources from University of Tokyo, LANL, Lawrence Livermore National Laboratory, Beijing and University of Zürich to give negative values of $m^2(v_e)$ which were not of physical relevance. To deal with these controversial results a new generation of spectrometers, "MAC-E Filters" were introduced at Mainz (1991) and Troitsk (1994) experiments. These spectrometers had higher resolution and luminosity as compared to the previous generation spectrometers. Even these experiments agreed with the negative values of $m^2(v_e)$ in their initial runs, but with further analysis and improvement in their experiments, they provided a non-negative value. The Mainz group eliminated their negative values by taking into account the energy losses due to the large inelastic scattering obtained by the inhomogeneous nature of their source material. The Triotsk group improved their results by incorporating the influence of large-angle scattering of electrons magnetically trapped in their tritium source. In their further runs, the Mainz and Triotsk experiments established an upper limit of 2.3 eV (95% C.L.) and 2.05 eV (95% C.L.) respectively on the $v_e$ mass [19,20].

Although, the Mainz and Triotsk experiment have provided ground breaking results on the upper bound of neutrino masses, still they have not been able to exploit the technology of MAC-E Filters to its limits. KATRIN has been designed to serve the purpose of exploiting the MAC-E Filter to its technological limits by increasing the sensitivity of the experiment to the sub eV regime. KATRIN combines the windowless gaseous molecular tritium source (WGTS) used in the LANL experiment with a spectrometer based on the principle of magnetic adiabatic collimation with electrostatic filtering (MAC-E Filter) to investigate the end point region of Tritium β decay spectrum with large statistics and extremely high resolution. Reducing the systematic errors by two orders of magnitude and enhancing the sensitivity on the neutrino mass by one order of magnitude, KATRIN has been able to obtain an upper limit of 1.1 eV on the $m(\nu_e)$ (90% C.L.) after four weeks of initial run in 2019. Fig.2 explains the scatter plot of fit values for the $m^2(\nu_e)$ and the effective β decay endpoint $E_0$ obtained from the KATRIN experiment [21].

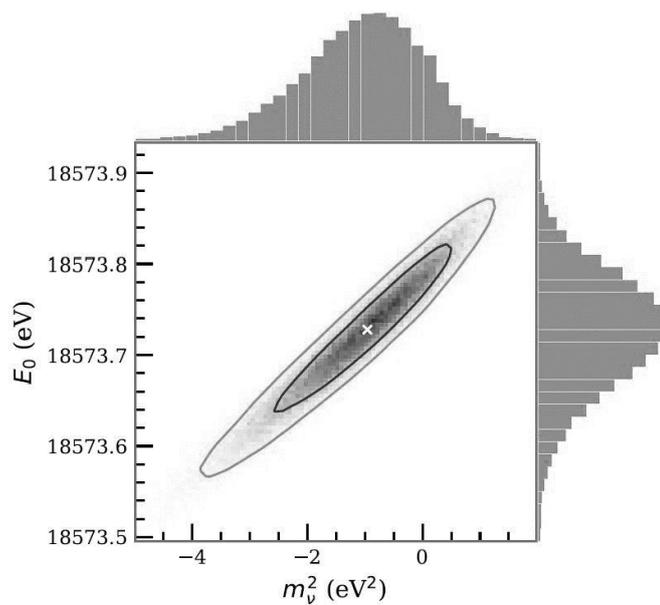

Fig. 2

With exponential improvements in a month of initial run over the previous results, KATRIN will be able to achieve a sensitivity of 0.2 eV after five years of data collection. Further future experiments like Project 8, ECHo, MARE and PTOLEMY are expected to provide neutrino mass limits with sub eV precision in the near future. Electron Capture $^{163}$Ho experiment (ECHo) and Microcalorimeter Arrays for a Rhenium Experiment (MARE) will use the concept of calorimetry, while Project 8 will be based on the detection of decay electron by the method of cyclotron radiation emission in a magnetic field. PTOLEMY will use neutrino capture on tritium as a function of the neutrino mass scale to produce a scalable model of Cosmic Neutrino Background (CNB). Further a search for $0\nu\beta\beta$ will be able to emphasize about the Majorana nature of neutrinos and provide information about the absolute mass scale of neutrinos. The results from these experiments in the near future will provide an explanation on the degeneracy of neutrino masses and its importance in cosmological studies.